\documentclass[epj, a4paper]{svjour}

\usepackage{graphicx}
\usepackage{bm,bbm}
\usepackage{amssymb,amsfonts,amsmath}
\usepackage[english]{babel}
\usepackage[nolist]{acronym}
\usepackage{math4phy}
\usepackage{xcolor}
\usepackage{hyperref}

\begin{document}

\title{Normal form expansions and thermal decay rates of Bose-Einstein 
       condensates with short- and long-range interaction}
\mail{andrej.junginger@itp1.uni-stuttgart.de}
\author{Andrej Junginger \and Teresa Schaller \and Gela H\"ammerling \and 
J\"org Main \and G\"unter Wunner}
\institute{Institut f\"{u}r Theoretische Physik 1, 
             Universit\"{a}t Stuttgart, 
             70550 Stuttgart, Germany}
\date{\today}

\authorrunning{A.~Junginger \etal}
\titlerunning{Normal form expansions and thermal decay rates of Bose-Einstein 
       condensates \ldots}

\newcommand{\ie}{i.\,e.}
\newcommand{\eg}{e.\,g.}
\newcommand{\cf}{cf.}
\newcommand{\etal}{\textsl{et~al.}}
\newcommand{\FIG}{Fig.}
\newcommand{\FIGS}{Figs.}
\newcommand{\SEC}{Sec.} \newcommand{\SECS}{Secs.}
\newcommand{\EQ}{Eq.}
\newcommand{\EQS}{Eqs.}
\newcommand{\REF}{Ref.}
\newcommand{\REFS}{Refs.}
\newcommand{\BEC}{BEC} \newcommand{\BECs}{BECs}
\renewcommand{\vec}[1]{\boldsymbol{#1}}
\renewcommand{\aa}{\vec{a}} 
\newcommand{\bb}{\vec{b}} \newcommand{\BB}{\vec{B}}
\newcommand{\cc}{\vec{c}} \newcommand{\CC}{\vec{C}}
\ifdefined\ee
  \renewcommand{\ee}{\vec{e}} \newcommand{\EE}{\vec{E}}
\else
  \newcommand{\ee}{\vec{e}} \newcommand{\EE}{\vec{E}}
\fi
\newcommand{\ff}{\vec{f}} \newcommand{\FF}{\vec{F}}
\newcommand{\hh}{\vec{h}} \newcommand{\HH}{\vec{H}}
\newcommand{\ii}{\vec{i}} \newcommand{\II}{\vec{I}}
\newcommand{\jj}{\vec{j}} \newcommand{\JJ}{\vec{J}}
\newcommand{\kk}{\vec{k}} \newcommand{\KK}{\vec{K}}
\newcommand{\mm}{{\vec{m}}} \newcommand{\MM}{\vec{M}}
\newcommand{\nn}{\vec{n}} \newcommand{\NN}{\vec{N}}
\newcommand{\oo}{\vec{o}} \newcommand{\OO}{\vec{O}}
\newcommand{\pp}{\vec{p}} \newcommand{\PP}{\vec{P}}
\newcommand{\qq}{\vec{q}} \newcommand{\QQ}{\vec{Q}}
\newcommand{\rr}{\vec{r}} \newcommand{\RR}{\vec{R}}
\newcommand{\vu}{\vec{u}} \newcommand{\UU}{\vec{U}}
\newcommand{\vv}{\vec{v}} \newcommand{\VV}{\vec{V}}
\newcommand{\xx}{\vec{x}} \newcommand{\XX}{\vec{X}}
\newcommand{\yy}{\vec{y}} \newcommand{\YY}{\vec{Y}}
\newcommand{\zz}{\vec{z}} \newcommand{\ZZ}{\vec{Z}}
\newcommand{\ud}{\mathrm{d}} \newcommand{\uD}{\mathrm{D}}
\newcommand{\ue}{\mathrm{e}} \newcommand{\uE}{\mathrm{E}}
\newcommand{\ui}{\mathrm{i}} \newcommand{\uI}{\mathrm{I}}
\newcommand{\re}{\mathrm{r}}
\newcommand{\im}{\mathrm{i}}
\newcommand{\Nbbm}{\mathbbm{N}}
\newcommand{\Qbbm}{\mathbbm{Q}}
\newcommand{\Rbbm}{\mathbbm{R}}
\newcommand{\Zbbm}{\mathbbm{Z}}
\newcommand{\Cbbm}{\mathbbm{C}}
\ifdefined\abs
  \renewcommand{\abs}[1]{\left|#1\right|}
\else
  \newcommand{\abs}[1]{\left|#1\right|}
\fi
\newcommand{\p}{\partial}
\newcommand{\Ng}{{N_{\!\text{g}}}}
\newcommand{\Tc}{T_\text{c}}
\newcommand{\kB}{k_\text{B}}
\newcommand{\transpose}{\mathsf{T}}
\newcommand{\dof}{{{d}}}
\newcommand{\drrr}{\ud^3r}
\newcommand{\Vext}{V_\text{ext}}
\newcommand{\Vint}{V_\text{int}}
\newcommand{\Vc}{V_\text{c}}
\newcommand{\Vlr}{V_\text{lr}}
\newcommand{\fluc}{\hat\delta}
\newcommand{\acrit}{a_\text{crit}}
\newcommand{\add}{a_\text{dd}}
\newcommand{\Emf}{E_\text{mf}}
\renewcommand{\Im}{\mathrm{Im}}

\abstract{
The thermally induced coherent collapse of Bose-Einstein condensates at finite 
temperature is the dominant decay mechanism near the critical scattering length 
in condensates with at least partially attractive interaction. The collapse 
dynamics out of the ground state is mediated by a transition state whose 
properties determine the corresponding decay rate or lifetime of the condensate.
In this paper, we perform normal form expansions of the ground and the 
transition state of condensates with short-range scattering interaction as well 
as with anisotropic and long-range dipolar interaction in a variational 
framework. This method allows one to determine the local properties of these 
states, \ie\ their mean-field energy, their normal modes, the coupling between 
different modes, and the structure of the reaction channel to any desired order.
We discuss the physical interpretation of the transition state as a certain 
density distribution of the atomic cloud and the behavior of the single normal 
form contributions in dependence on the s-wave scattering length. Moreover, we 
investigate the convergence of the local normal form when using extended 
Gaussian variational approaches, and present the condensate's decay rate.
}



\maketitle

\section{Introduction}

Starting from their first experimental realization in 1995 
\cite{Anderson1995,Bradley1995,Davis1995}, the field of Bose-Einstein 
condensates (BECs) has grown rapidly and it is the subject of numerous 
experimental and theoretical investigations today. 
By far, most of today's experiments are performed on such macroscopic quantum 
objects made of alkali metals, in which the interaction between the single 
bosons is the short-range, spherically symmetric, low-energy s-wave scattering.
Beyond these, also BECs with an additional long-range and anisotropic 
dipole-dipole interaction (DDI) have been realized experimentally 
\cite{Griesmaier2005,Lu2011,Aikawa2012}, which are of great interest, because 
the interaction between two bosons depends on their relative orientation. As a 
consequence, interesting phenomena have been predicted in dipolar BECs, such as 
isotropic as well as anisotropic solitons 
\cite{Pedri2005,Nath2009,Tikhonenkov2008}, biconcave or structured ground state 
density distributions \cite{Dutta2007,Ronen2007,Goral2000}, stability diagrams 
that depend on the trap geometry \cite{Koch2008,Santos2000,Goral2002}, radial 
and angular rotons \cite{Ronen2007,Santos2003,Wilson2008}, and anisotropic 
collapse dynamics \cite{Metz2009,Lahaye2008}.

If the particle interaction between the single bosons in the condensate is (at 
least partially) attractive, its ground state is metastable and several 
mechanisms can contribute to the decay of the atomic cloud. These include \eg\ 
inelastic three-body collisions, macroscopic quantum tunneling 
\cite{Stoof1997,Marquardt2012}, the decrease of the s-wave scattering length 
below its critical value \cite{Ronen2007}, or dipolar relaxation 
\cite{Hensler2003}.

Another important decay mechanism is the thermally induced coherent collapse of 
the condensate 
\cite{Huepe1999,Huepe2003,Junginger2012a,Junginger2012b,Junginger2012d,Junginger2013b}. 
This process is based on the fact that quasi-particle excitations in a BEC at 
finite temperature $T>0$ lead to time-dependent density fluctuations of the gas. 
If the particle interaction is attractive, these fluctuations can induce the 
collapse of the condensate, when the density locally becomes high enough so that 
the attraction can no longer be compensated by the quantum pressure. This 
process is important near the critical scattering length where the attraction 
between the bosons becomes dominant, and it is mediated by a transition state. 
The latter is a collectively excited, stationary state of the condensate which 
typically exhibits a locally increased density in some region and which emerges 
together with the ground state in a tangent bifurcation at a critical value of 
the scattering length.
The thermally induced collapse dynamics is then of the type ``reactants $\to$ 
transition state $\to$ products'', so that the condensate's decay rate or 
lifetime, respectively, can be calculated by applying transition state theory 
(TST) \cite{Pechukas1981,Haenggi1990,Truhlar1996,Waalkens2008}:
With respect to the BEC's ground state the transition state forms an energy 
barrier that needs to be crossed in order to induce the collapse.  Finally, the 
height of the energy barrier together with the local properties of the ground 
and the transition state determine the reaction dynamics.

In this paper, we investigate the thermal decay of condensates at low 
temperatures $T$ in the region $0<T_0<T\ll\Tc$, \ie\ above a temperature 
$T_0\sim \hbar\omega/\kB$ where collective oscillations with frequency $\omega$ 
are thermally excited and significantly below the critical temperature $\Tc$ 
where the macroscopic occupation of the ground state sets in. In that 
temperature regime, the thermal excitations of the condensate are of collective 
nature and single-particle excitations can be neglected 
\cite{Junginger2012d,Junginger2013b}.
In addition, three-body collisions due to high condensate densities do not 
influence the decay rate corresponding to the thermally induced coherent 
collapse, because they only become important \emph{after} the transition state 
has been crossed.
Therefore, a suitable framework for the 
theoretical description of the BEC is that of the Gross-Pitaevskii equation 
(GPE).
Moreover, a variational approach to the GPE is especially appropriate to 
investigate the transition state, because in this framework, the transition 
state is a fixed point of the corresponding dynamical equations, and its local 
properties are determined by the latters' series expansion.
A particularly appropriate set of coordinates into which this expansion can be 
transformed are its normal form coordinates. These have the advantage that 
locally they can be chosen as classical canonical coordinates 
\cite{Junginger2014a}, in which the condensate's mean-field energy functional 
serves as a classical Hamilton function that fully describes the dynamics of the 
BEC. The complete information about the transition state can then be extracted 
from the expansion coefficients of this Hamiltonian.

In this paper, we focus on the normal form expansions of both the ground and the 
transition state of BECs with short-range interaction \cite{TeresaBachelor} as 
well as with long-range and anisotropic DDI \cite{GelaBachelor}. We present the 
physical interpretation of the transition state as a certain density 
distribution of the atomic cloud. Moreover, we discuss the behavior of the 
single expansion coefficients which describe the fixed point energies, the 
corresponding elementary excitations and the coupling between the different 
normal modes. 
Finally, we calculate the contribution of the thermally induced collapse to the 
decay rate of the condensate at experimentally relevant temperatures.

Our paper is organized as follows:
In \SEC~\ref{sec:theory} we discuss the theoretical description of the BEC in a 
variational framework, for which we introduce the GPE and a time-dependent 
variational principle. Moreover, we review the construction of the local normal 
form Hamiltonian and, with it, the calculation of the decay rate by applying 
TST.
In \SEC~\ref{sec:results}, we present the results for \BECs\ with short-range 
scattering interaction as well as with long-range and anisotropic DDI.

\section{Theory} \label{sec:theory}

\subsection{BECs at ultracold temperatures}

At ultra-cold temperatures, the dynamics of a BEC is determined by the GPE 
(units given below)
\begin{align}
\begin{split}
  \ui \pop{}{t} \psi(\rr,t) 
  =
  \hat H \psi(\rr,t) 
  =
  \Bigl( - \Delta + \Vext + \Vc + \Vlr \Bigr) \psi(\rr,t) \,.
\end{split}
\label{eq:GPE}
\end{align}
Here, $\Vext$ is an external trapping potential, the contact interaction $\Vc = 
8 \pi a N \abs{\psi(\rr,t)}^2$ describes low-energy collisions between the 
bosons via the s-wave scattering length $a$ and the particle number $N$, and 
$\Vlr$ is a possible long-range particle interaction.

In case of a BEC without long-range interaction ($\Vlr=0$) the internal symmetry 
of the system is spherical and we, therefore, also choose a spherically 
symmetrical external trapping potential $\Vext=\frac{m}{2}\omega^2\rr^2$.
The form of the GPE \eqref{eq:GPE} with the given interaction potentials is then 
obtained by using the oscillator length $r_0=\sqrt{\hbar/m\omega}$ as a natural 
unit of length, with $m$ being the mass of the bosons and $\omega$ being the 
trap frequency. Natural energy and time scales are then given by 
$E_0=\hbar\omega/2$ and $t_0=\hbar/E_0$. Furthermore, we use $m_0=2m$ as a unit 
of mass.

In case of a BEC with long-range and anisotropic DDI in which all dipoles are 
aligned in $z$-direction by an external magnetic field, the long-range part of 
the interaction potential in the GPE reads
\begin{align}
  \Vlr  &= \add N
    \int \drrr' ~ \frac{1-3\cos^2\theta}{\abs{\rr-\rr'}^3} 
    \abs{\psi(\rr',t)}^2 \,.
\label{eq:potentials-DDI}%
\end{align}
The alignment of the dipoles naturally induces a cylindrical symmetry to the BEC 
and we, therefore, adapt the symmetry of the external trap to 
$\Vext=\rho^2+\lambda^2z^2$ where $\rho^2=x^2+y^2$.
As a length scale we use the radial oscillator length 
$r_0=\sqrt{\hbar/m\omega_\rho}$, and we define the trap strength in 
$z$-direction via the trap aspect ratio $\lambda=\omega_z/\omega_\rho$. In these 
units, the strength of the DDI reads $\add=\mu_0\mu^2m/(2\pi\hbar^2r_0)$ with 
$\mu$ being the magnetic moment of the atoms.

\subsection{Time-dependent variational approach}

Common methods to solve the GPE \eqref{eq:GPE} are \eg\ its direct numerical 
integration or the discretization of the wave function on grids. The 
condensate's dynamics and its ground state can then be calculated by applying 
the split-operator method and an imaginary-time evolution.
As already mentioned above, a more suitable framework for the purposes of this 
paper is the description within a variational approach. Therein, the GPE is 
solved approximately by replacing the original wave function $\psi(\rr,t)$ by a 
trial wave function
\begin{equation}
   \psi(\rr,t) = \psi(\rr, \zz(t)) \,. 
   \label{eq:TDVP-trial-wave-function}
\end{equation}
Here, $\zz(t) = [z_1(t), z_2(t), \ldots, z_\dof(t)]^\transpose \in \Cbbm^\dof$ 
is a set of $\dof$ complex and time-dependent variational parameters, and the 
time evolution of the wave function is completely determined by them.
In the framework of the variational approach, the mean-field energy functional 
of the system is given by the expectation value of the Hamilton operator
\begin{equation}
  E(\zz) = 
  \bra[Big]{\psi(\rr,\zz)} 
    -\Delta + \Vext + \frac{1}{2} ( \Vc + \Vlr )
  \ket[Big]{\psi(\rr,\zz)} \,,
  \label{eq:TDVP-energy-functional}
\end{equation}
where the factor $1/2$ is included to avoid a double-counting of the 
two-particle interactions.
In order to describe the dynamics of the system in the Hilbert subspace which is 
spanned by the variational ansatz \eqref{eq:TDVP-trial-wave-function}, we apply 
the Dirac-Frenkel-McLachlan variational principle 
\cite{Frenkel1934,McLachlan1964}. This requires minimizing the norm of the 
difference between the left- and the right-hand side of the GPE \eqref{eq:GPE},
\begin{equation}
   I 
   = 
   \norm{ \ui \phi - \hat{H} \psi }^2
   \stackrel{!}{=} \text{min.} 
   \label{eq:McLachlan-variational-principle}
\end{equation}
where, the arguments of the wave function $\psi$ have been omitted for brevity. 
The quantity $I$ is minimized with respect to $\phi$ and $\phi=\dot\psi$ is set 
afterwards which means that the GPE is solved within the Hilbert subspace of the 
variational ansatz with the least possible error.
Proceeding from the complex variational parameters $\zz$ to their real and 
imaginary parts $\xx=(\zz^\re, \zz^\im)^\transpose \in \Rbbm^{2\dof}$, it was 
shown in \REF\ \cite{Junginger2014a} that minimizing 
\EQ~\eqref{eq:McLachlan-variational-principle} leads to the noncanonical 
Hamiltonian equations of motion
\begin{equation}
  K(\xx) \, \dot{\xx}
  = - \pop{E(\xx)}{\xx} \,,
  \label{eq:TDVP-equations-of-motion}
\end{equation}
where $K_{mn} = 2 \Im \braket { \pop{\psi}{x_m} } { \pop{\psi}{x_n} }$ and 
$E(\xx)=E(\zz) \left.\right|_{\zz\to\xx}$.

In order to apply the variational approach to a BEC in a harmonic trap, a 
natural choice for the trial wave function \eqref{eq:TDVP-trial-wave-function} 
is a Gaussian one. Deviations from the pure Gaussian form which occur due to 
particle interactions can then be taken into account by using \emph{coupled} 
Gaussian trial wave functions
\begin{equation}
  \psi(\rr, \zz) = \sum_{k=1}^\Ng g_k(\rr, \zz) \,,
  \label{eq:TDVP-coupled-Gaussian}
\end{equation}
where we have omitted the explicit time-dependence of the variational parameters 
$\zz$ for brevity.
Depending on the inherent symmetry of the physical system, we will choose one of 
the following forms in this paper:
\begin{subequations}
\begin{align}
  g_k(\rr, \zz) &= \exp (A_r^k r^2 + \gamma^k) \,, 
  \label{eq:TDVP-Gaussian-radial}\\
  g_k(\rr, \zz) &= \exp (A_\rho^k \rho^2 + A_z^k z^2 + \gamma^k) \,.
  \label{eq:TDVP-Gaussian-cyl}
\end{align}
\end{subequations}
Here, we use the complex variational parameters 
$\zz=(A^k_r,A^k_\rho,A^k_z,\gamma^k)^\transpose$: the parameters 
$A^k_{r,\rho,z}$ determine the width of each Gaussian and $\gamma^k$ are the 
norm and phase, respectively. 
Equation \eqref{eq:TDVP-Gaussian-radial} is an appropriate choice for the 
radially symmetrical system without DDI and \EQ~\eqref{eq:TDVP-Gaussian-cyl} for 
the dipolar system with cylindrical symmetry. We note that, because the total 
wave function is normalized to $\| \psi(\rr,t) \|^2=1$ and its global phase is 
free, the total number of independent variational parameters is reduced by one. 
Consequently, there remain $\dof=2\Ng-1$ degrees of freedom in case of the 
ansatz \eqref{eq:TDVP-Gaussian-radial} and $\dof=3\Ng-1$ in case of 
\EQ~\eqref{eq:TDVP-Gaussian-cyl}.
For the detailed application of the time-dependent variational approach, \ie\ 
the evaluation of the energy functional \eqref{eq:TDVP-energy-functional} and 
the dynamical equations \eqref{eq:TDVP-equations-of-motion}, we refer the reader 
to \REF\ \cite{Rau2010a}.

\subsection{Normal form expansion and TST}
\label{sec:theory-NF}

In the investigations of BECs, fixed points of the dynamical equations 
\eqref{eq:TDVP-equations-of-motion} are of special interest, because they 
correspond to stationary states of the system as, \eg, its ground or transition 
state.
Just as important as the fixed points themselves are their local properties, 
since these determine the elementary excitations and the structure of the 
reaction channel.
Within the lowest-order approximation to the condensate's dynamics, a possible 
approach to determine the elementary excitations are the Bogoliubov-de Gennes 
equations. The latter result from a linearization of the GPE \eqref{eq:GPE} for 
small deviations from its ground state and yield the BEC's collective 
frequencies $\pm \omega_i$. 
The same can be obtained from the variational approach by linearizing the 
dynamical equations \eqref{eq:TDVP-equations-of-motion} at a fixed point giving 
the finite set of frequencies $(\pm\omega_1, \pm\omega_2, \ldots, \pm\omega_d)$ 
\cite{Kreibich2012,Kreibich2013a}.

Beyond the linear approximation of the dynamics, we are, in this paper, also 
interested in the higher-order corrections. A systematic way to investigate the 
local fixed point properties to any desired order is a normal form expansion of 
the dynamical equations \eqref{eq:TDVP-equations-of-motion}. This method has 
been described in detail in \REF\ \cite{Junginger2014a} and we will only discuss 
it here very briefly:
The essential procedure to bring the noncanonical Hamiltonian system into its 
canonical normal form is based on a power series expansion of the energy 
functional \eqref{eq:TDVP-energy-functional} and the dynamical equations 
\eqref{eq:TDVP-equations-of-motion} at the respective fixed point. Successive 
Lie transforms which are performed order by order bring the system into its 
normal form.
Furthermore, the corresponding generating function of the transformation is 
chosen in such a way that the resulting dynamical equations as well as the 
energy functional fulfill canonical equations, \ie\ the normal form coordinates 
are canonical ones. 
Finally, the energy functional \eqref{eq:TDVP-energy-functional} in these 
coordinates serves as a classical Hamilton function. 
Using the multi-index notation $\xx^\mm = x_1^{m_1} x_2^{m_2}\ldots 
x_\dof^{m_\dof}$ and $\abs{\mm}=m_1+m_2+\ldots+m_\dof$, the latter is a 
multivariate polynomial in action variables $\JJ$, 
\begin{equation}
  H(\JJ) = \sum_{\abs{\mm}=0}^\chi \xi_\mm \JJ^\mm \,,
  \label{eq:Ham-NF}
\end{equation}
where $\chi$ is the normal form order chosen and $\xi_\mm$ are the coefficients 
of the expansion. The zeroth-order coefficient $\xi_{\abs{\mm}=0} = \Emf$ is the 
mean-field energy and those of first-order are the oscillation frequencies 
$\xi_{\abs{\mm}=1} = \omega_i$ ($i=1,\ldots,\dof$). The polynomial structure of 
\EQ~\eqref{eq:Ham-NF} can always be obtained if the first-order coefficients of 
the expansion are rationally independent, \ie\ the equation
\begin{equation}
  \sum_{\mm} n_\mm \, \xi_{\abs{\mm}=1} = 0
  \label{eq:rational-independent}
\end{equation}
with $n_\mm \in \Zbbm$ has only the trivial solution $n_\mm=0$.
It is emphasized that the expansion coefficients $\xi_{\mm}$ in 
\EQ~\eqref{eq:Ham-NF} contain the full information about the local dynamics at 
the fixed points.

In addition to the simple structure of \EQ~\eqref{eq:Ham-NF} in normal form 
coordinates, a big advantage of this canonical Hamiltonian is that it allows one 
to apply all methods which are known from classical Hamiltonian mechanics. One 
important application is the field of TST 
\cite{Pechukas1981,Haenggi1990,Truhlar1996,Waalkens2008} where reactions are 
described  qualitatively and quantitatively that are mediated by a transition 
state in phase space: If $J_1$ is the reaction coordinate, then the transition 
state is defined by $J_1=0$ and the thermally averaged reaction rate is 
\cite{Haenggi1990}
\begin{equation}
  \Gamma =
  \frac{1}{2\pi \beta}
  \frac{
    \int \ud J_2 \ldots \ud J_\dof \, 
    \exp \bigl(-\beta H(0, J_2, \ldots, J_\dof)  \bigr) 
  }{
    \int \ud J'_1 \ldots \ud J'_\dof \, 
    \exp \bigl(-\beta H'(J'_1, \ldots, J'_\dof) \bigr) 
  }\,.
\label{eq:TST-phase-space-Gamma}
\end{equation}
Here, $H$ is the normal form Hamiltonian of the system at its transition state, 
$H'$ is the corresponding one at the ground state, and $\beta = 1/\kB T$ is the 
inverse temperature.

\section{Application to BECs with short- and long-range interaction}
\label{sec:results}

\subsection{BECs with short-range interaction}
\label{sec:alkali}

A BEC with short-range interaction is described by the GPE \eqref{eq:GPE} with 
$\Vlr=0$ and, as already mentioned above, we focus on a system that is confined 
in a spherically symmetrical trap $\Vext=\rr^2$ for simplicity.
Searching for fixed points of the corresponding dynamical equations 
\eqref{eq:TDVP-equations-of-motion} in the variational framework, one can find 
two stationary states above a critical scattering length $\acrit$, one of which 
corresponds to the ground state and the other is its transition state.

\begin{figure}[t]
\includegraphics[width=\columnwidth]{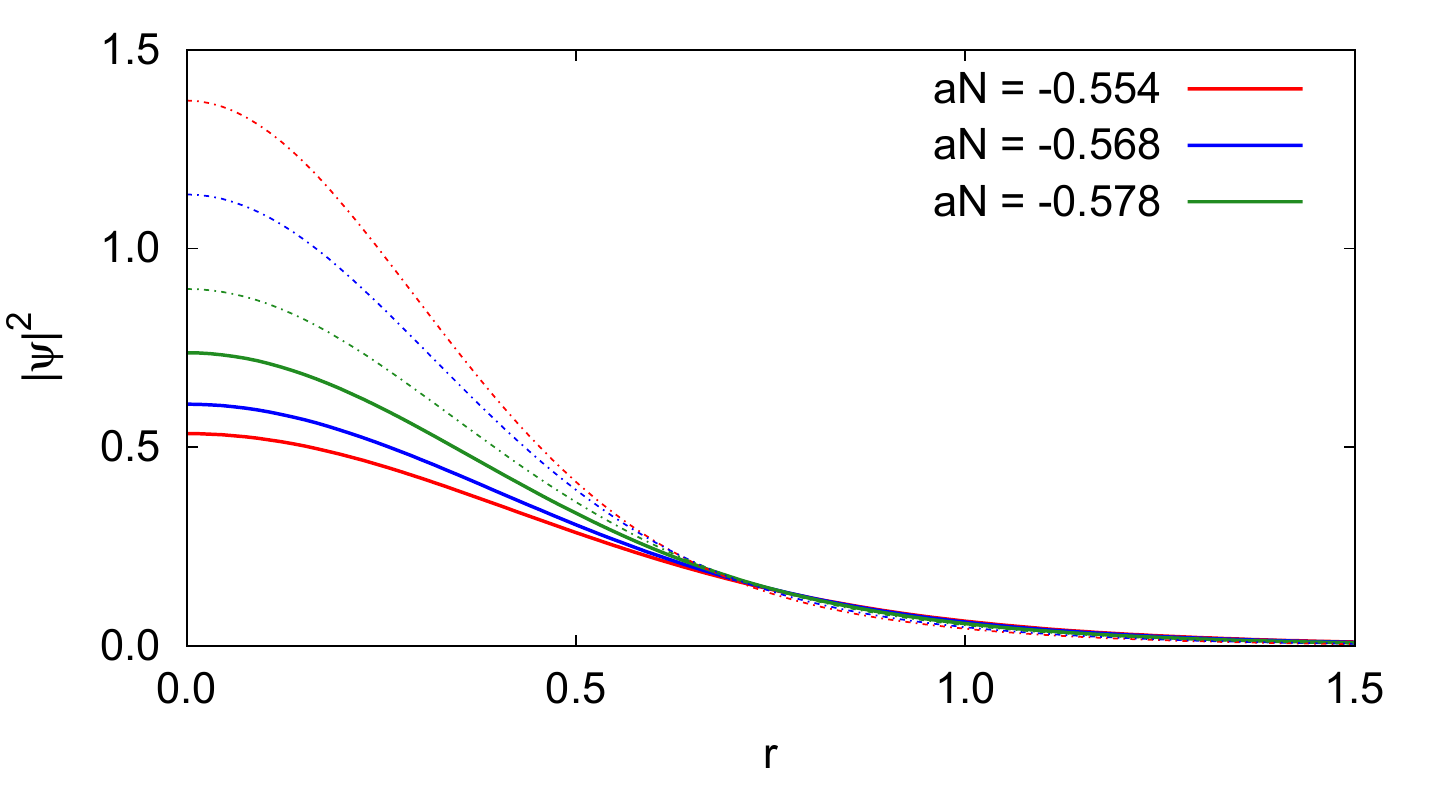}
\caption{
(Color online)
Density distribution $\abs{\psi(\rr)}^2$ of the condensate in its ground (solid 
lines) and transition state (dash-dotted lines) at different values of the 
s-wave scattering length and for $\Ng=2$ coupled Gaussians.
Compared to the ground state, the transition state exhibits a higher density at 
the center of the trap and a lower density far away from it.
With decreasing scattering length, the maximum density of the ground state 
increases, while that of the transition state decreases, and at the critical 
scattering length $\acrit N=-0.579$ they match.
}
\label{fig:alkali-wave-functions}
\end{figure}

In \FIG~\ref{fig:alkali-wave-functions}, the density distributions  
$\abs{\psi(\rr)}^2$ of the ground (solid lines) and the transition state 
(dash-dotted lines) are shown for different values of the s-wave scattering 
length and $\Ng=2$ coupled Gaussians.
Compared to the ground state, the transition state in general exhibits a higher 
density at the center of the trap and a lower density far away from it. 
With decreasing scattering length, the density of the ground state increases, 
while that of the transition state decreases, and at the critical value $\acrit 
N$ of the scattering length, the two states become identical.
Because the interaction is attractive ($aN<0$) and the density distribution 
directly enters the contribution of the particle scattering, the transition 
state physically represents a highly attracting configuration of the condensate. 
More precisely, its physical interpretation is a density distribution on the 
edge of the BEC's collapse: For any higher (local) density, the attractive 
interaction would dominate the quantum pressure leading to the collapse of the 
condensate. Any lower density would result in an excited but stable BEC. This 
interpretation can be verified by actually calculating the dynamics of the BEC 
\cite{Junginger2013b,Cartarius2008a}, which reveals its collapse in the center 
of the trap after the transition state has been crossed.

\begin{figure}[t]
\includegraphics[width=\columnwidth]{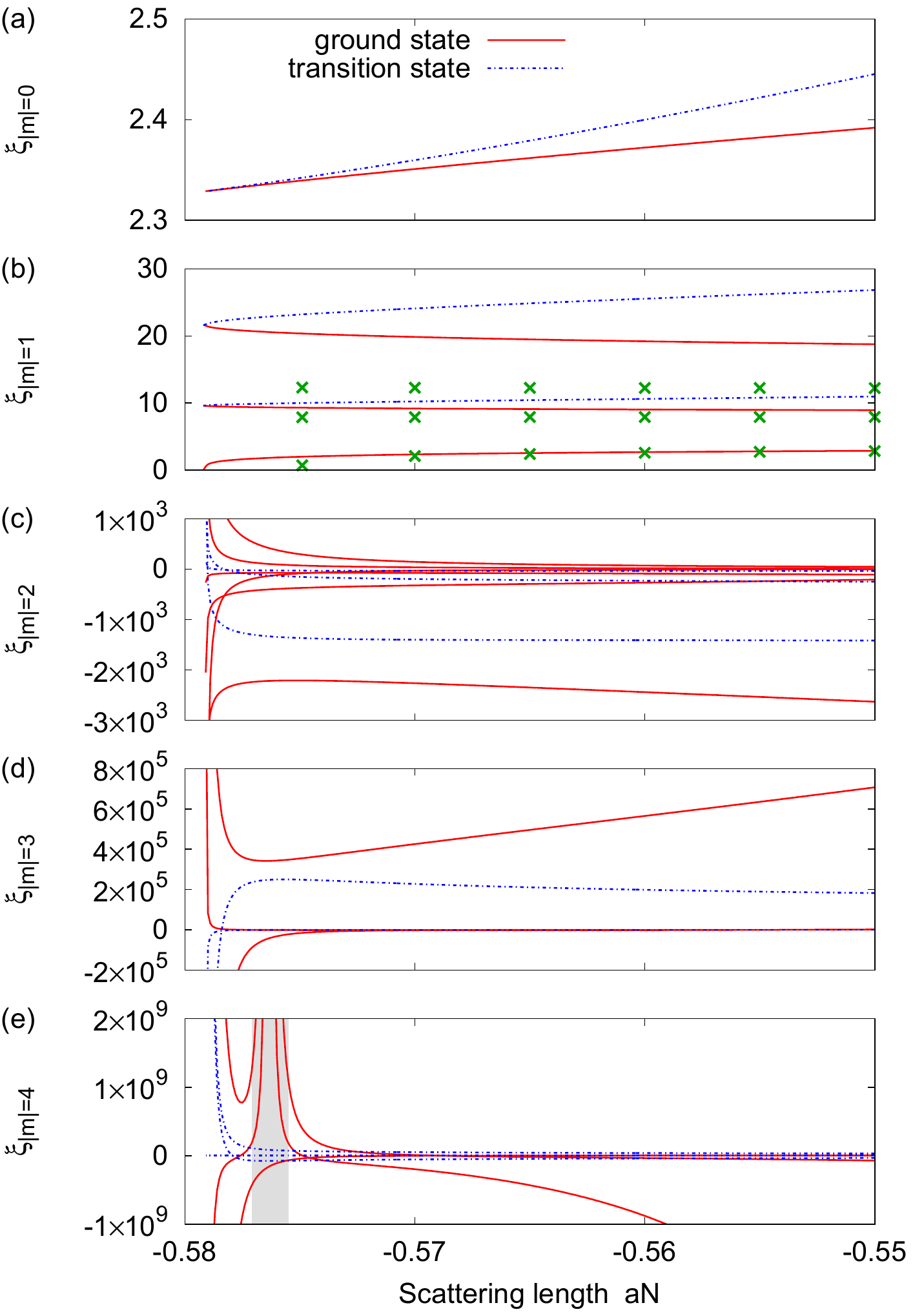}
\caption{
(Color online)
Coefficients of the normal form expansion in different orders for an 
approximation of the condensate wave function with $\Ng=2$ coupled Gaussian wave 
functions.
(a) The zeroth-order coefficient is the mean-field energy ($\xi_{\abs{\mm}=0} = 
\Emf$) of the stationary state.
(b) The first-order coefficients correspond to the eigenvalues of the linearized 
dynamical equations ($\xi_{\abs{\mm}=1} = \omega_i$, $i=1,\ldots,\dof$) and 
describe the frequencies of the Bogoliubov quasi-particle modes.
For comparison numerically exact results (taken from \REF\ \cite{Kreibich2012}) 
are shown as green crosses. 
(c)--(e): Higher-order terms give corrections of the dynamics in the vicinity of 
the fixed points and they define the coupling strength of the single 
quasi-particle modes (only a selection is presented for the sake of clarity). 
The gray bar in (e) is intended to highlight the divergence of some coefficients 
(\cf\ \FIG~\ref{fig:alk-resonance}).
}
\label{fig:alkali-NF-coef}
\end{figure}

In \FIG~\ref{fig:alkali-NF-coef} the normal form expansion coefficients which 
fully describe the local properties of the fixed points are shown for the ground 
and the transition state:
The zeroth-order coefficients $\xi_{\abs{\mm}=0} = \Emf$ [see 
\FIG~\ref{fig:alkali-NF-coef}(a)] are the mean-field energies of the stationary 
states. The energetically lower one is the metastable ground state of the BEC 
and the other excited state is the transition state. For $\Ng=2$, both these 
states emerge together in a tangent bifurcation at the critical value $ \acrit N 
\approx -0.579$ of the scattering length, below which the condensate no longer 
exists.
Because the transition state has a higher energy than the ground state, an 
energy barrier has to be crossed in order to induce the BEC's collapse. The 
height of this barrier is given by the energy difference between the two states 
and it is high for large values of the scattering length. By contrast, it 
decreases when one approaches the critical value and vanishes there.
Figure \ref{fig:alkali-NF-coef}(b) shows the first-order coefficients 
($\xi_{\abs{\mm}=1} = \omega_i$, $i=1,\ldots,\dof$) of the local normal form 
which correspond to the frequencies of the Bogoliubov quasi-particle modes.
All these coefficients show a smooth dependence on the scattering length. At the 
latter's critical value, two of them merge in each case. 

For comparison the three lowest numerically exact Bogoliubov frequencies are 
shown for some values of the scattering length (crosses; taken from \REF\ 
\cite{Kreibich2012}): One can see that the lowest first-order normal form 
coefficient quantitatively agrees very well with the lowest Bogoliubov mode. 
Also the second one is in good agreement with the numerical results while one 
observes deviations for the third mode. 
It has been shown by Kreibich \etal\ \cite{Kreibich2012,Kreibich2013a} that the 
low-energy modes are in general reproduced very well by the variational 
approach. The high-energy modes are more difficult to reproduce, but the 
convergence can be improved by using higher numbers of coupled Gaussians. 
We refer the reader to these references for a detailed discussion about the 
convergence behavior of the oscillation frequencies for different particle 
interactions and trapping potentials.

For larger deviations of the system from one of the fixed points, higher-order 
contributions become important which correspond to higher-order coupling 
coefficients between the different Bogoliubov modes. In the normal form 
approach, these coupling coefficients are the expansion coefficients of higher 
than the linear order and they are shown in 
\FIGS~\ref{fig:alkali-NF-coef}(c)--(e).
All three figures show that the higher-order corrections have quite large 
numerical values (on the order of up to $10^{9}$) as compared to the oscillation 
frequencies (on the order of $10^1$), and that they diverge at the critical 
scattering length $\acrit N$. This gives rise to the expectation that the 
corrections can have significant influence on the reaction rate, because their 
contribution becomes important in the vicinity of the critical scattering 
length.
Because of their physical meaning as coupling coefficients between the 
Bogoliubov modes, a possibility to check their numerical values would be to set 
up nonlinear and higher-order approximations to the GPE (\ie\ higher than the 
linearization leading to the Bogoliubov-de Gennes equations) and to determine 
the coupling order by order. This is, however, beyond the scope of this paper.

\begin{figure}[t]
\includegraphics[width=\columnwidth]{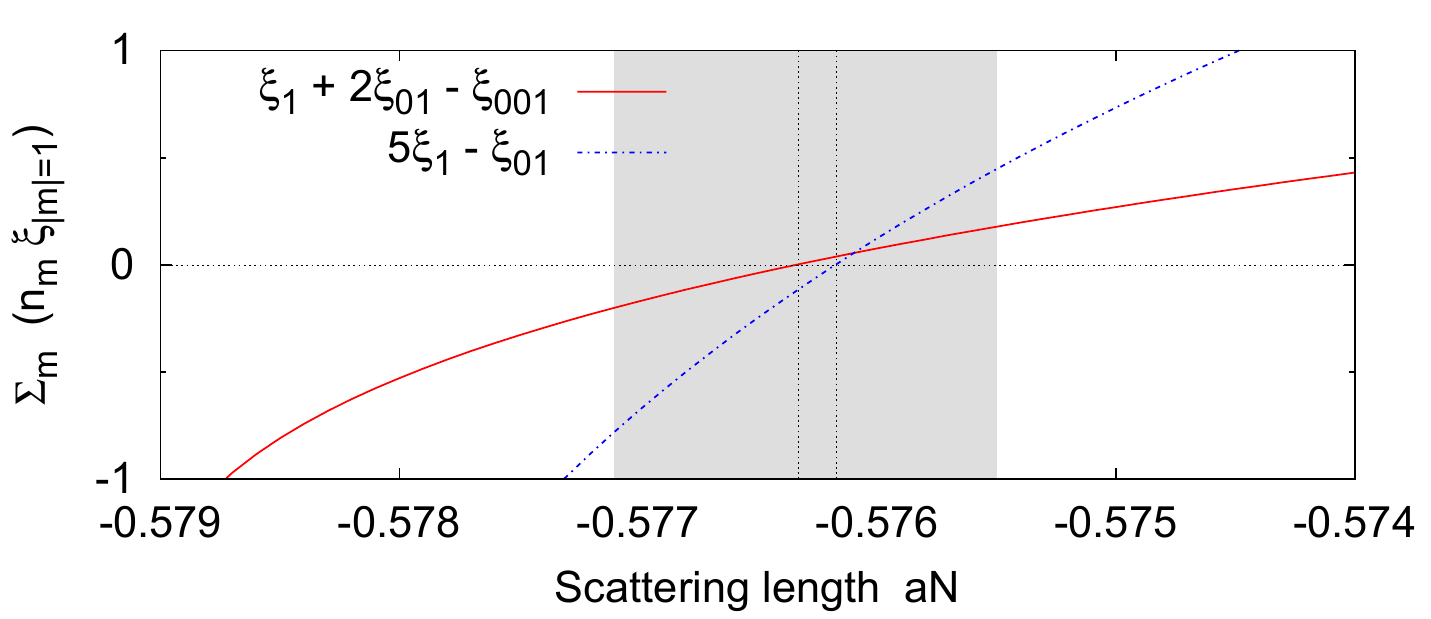}
\caption{%
(Color online)
Rational independence of the linear coefficients $\xi_{\abs{\mm}=1}$ for the 
same parameters as used in \FIG~\ref{fig:alkali-NF-coef}.
Equation \eqref{eq:rational-independent} is violated at $aN=-0.57633$ where 
$\xi_1 + 2\xi_{01} - \xi_{001} = 0$ and at $aN=-0.57617$ where $5\xi_1 - 
\xi_{01} = 0$. 
The width of the gray area in the background is the same as in 
\FIG~\ref{fig:alkali-NF-coef}(e).
}
\label{fig:alk-resonance}
\end{figure}

We note that the pole occurring in some fourth-order coupling terms  
[highlighted by the gray background in \FIG~\ref{fig:alkali-NF-coef}(e)] is a 
resonance in the normal form procedure. 
As shown in \FIG~\ref{fig:alk-resonance}, there are two nearby values of the 
scattering length $aN=-0.57633$ and $aN=-0.57617$, where the frequencies become 
rationally dependent, \ie\ \EQ~\eqref{eq:rational-independent} is violated. 
In this case there is a strong mode coupling of the condensate's higher 
harmonics in the respective order, which leads to the failure of the normal form 
procedure as described in \REF\ \cite{Junginger2014a}.

\begin{figure}[t]
\includegraphics[width=\columnwidth]{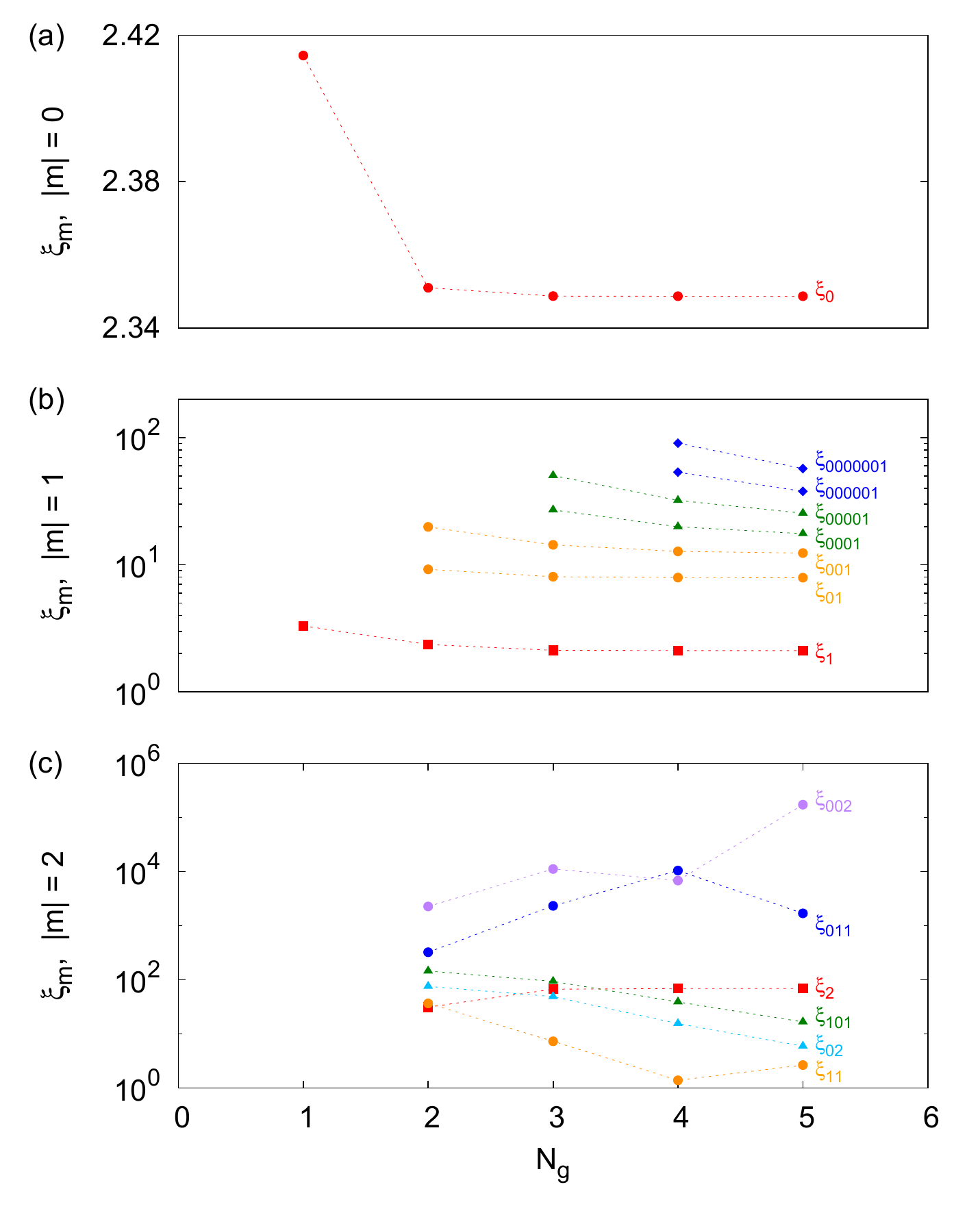}
\caption{
(Color online)
Behavior of the normal form expansion coefficients for a trial wave function in 
\EQ~\eqref{eq:TDVP-trial-wave-function} consisting of a different number $\Ng$ 
of coupled Gaussians. The scattering length is chosen as $aN=-0.57$.
In (a), the zeroth-order coefficient (fixed point energy) converges very fast.
The first-order contributions (Bogoliubov quasi-particle frequencies) in (b) 
show a monotonic decrease. The lowest terms converge fast while the higher terms 
are not yet converged for $\Ng=5$.
In the higher-order expansion coefficients (c), one observes the convergence of 
some coupling terms, while some others are not converged yet.
[Note that only a selection of coupling terms with a typical behavior is 
presented in (c) due to the huge amount of such terms.]
}
\label{fig:alkali-convergence}
\end{figure}


According to the variational ansatz \eqref{eq:TDVP-trial-wave-function}, where 
the number $\Ng$ of coupled wave functions appears as a free parameter, one 
expects that all the results depend on this parameter. Therefore, an important 
topic is the convergence of the normal form when the number of coupled Gaussians 
is varied.
For the following investigations of the single expansion coefficients, we will 
use a simplified notation in which each $\mm$-index in \EQ~\eqref{eq:Ham-NF} is 
only displayed up to its last nonzero entry and successive zeros are neglected 
(\eg\ the expansion coefficient $\xi_{021000}$ will be displayed as 
$\xi_{021}$). This makes it easy to compare expansion coefficients which result 
from different variational approaches in which the dimension of $\mm$ depends on 
the number $\Ng$ of coupled Gaussian wave functions.

In \FIG~\ref{fig:alkali-convergence}, we present the convergence behavior of a 
selection of coefficients in dependence on the number $\Ng$ of coupled Gaussian 
trial wave functions.
The zeroth-order coefficient $\xi_0$ in \FIG~\ref{fig:alkali-convergence}(a), 
\ie\ the fixed point energy, converges very fast. The most significant 
correction is observed when one increases from $\Ng=1$ to $\Ng=2$. In the last 
step shown ($\Ng=4$ to $\Ng=5$) the relative correction is about 
$4.4\times10^{-7}$, so that this value can be treated as converged.
The analogous behavior of the first-order coefficients is shown in 
\FIG~\ref{fig:alkali-convergence}(b). As already discussed in 
\SEC~\ref{sec:theory-NF}, the number of degrees of freedom is $\dof=2\Ng-1$. 
Therefore, two more terms occur with each increase of $\Ng$ which are indicated 
by the same symbols in this figure.
For a small number of coupled Gaussian wave functions, the corresponding 
coefficients are small and they correspond to low-frequency oscillation modes 
(\eg\ the lowest coefficient is the frequency of the BEC's breathing mode). The 
lowest terms only change marginally when the number $\Ng$ of coupled Gaussians 
is increased, thus convergence is observed early.
For the terms that correspond to more complicated higher-frequency oscillation 
modes, the corrections become larger, and they are still significant for 
$\Ng=5$. In this case, even more advanced trial wave functions will be required 
to observe convergence.
It is obvious throughout that the single coefficients decrease monotonically 
with larger values $\Ng$, so that one always expects the numerical results to 
overestimate the true values.

Finally, we present in \FIG~\ref{fig:alkali-convergence}(c) the behavior of a 
selection of second-order normal form contributions which correspond to the 
lowest-order coupling terms of the condensate oscillation modes:
It can be seen that the convergence of these coupling terms is not as simple as 
that of the zeroth- and first-order terms. The red squares in 
\FIG~\ref{fig:alkali-convergence}(c) exemplarily show a converging coupling 
coefficient. However, one can also observe other coefficients which exhibit 
throughout a monotonic decrease (triangles) or a nonmonotonic behavior (dots) 
and which are not yet converged.

As already mentioned above, the knowledge of the ground and the transition state 
of the BEC as well as their local properties allows one to calculate the 
condensate's thermal decay rate by applying TST. Therefore, a normal form 
expansion is performed at the ground and the transition state of the condensate 
which yields two normal form Hamiltonians $H$ and $H'$. The decay rate is, 
finally, given by \EQ~\eqref{eq:TST-phase-space-Gamma} where the integrals are 
evaluated numerically via a Monte Carlo integration. In the following, we 
present results for an exemplary BEC of $N=1000$ $^{87}$Rb atoms in a trap of 
strength $\omega=2\pi\times100\,$Hz.
%
\begin{figure}[t]
\includegraphics[width=\columnwidth]{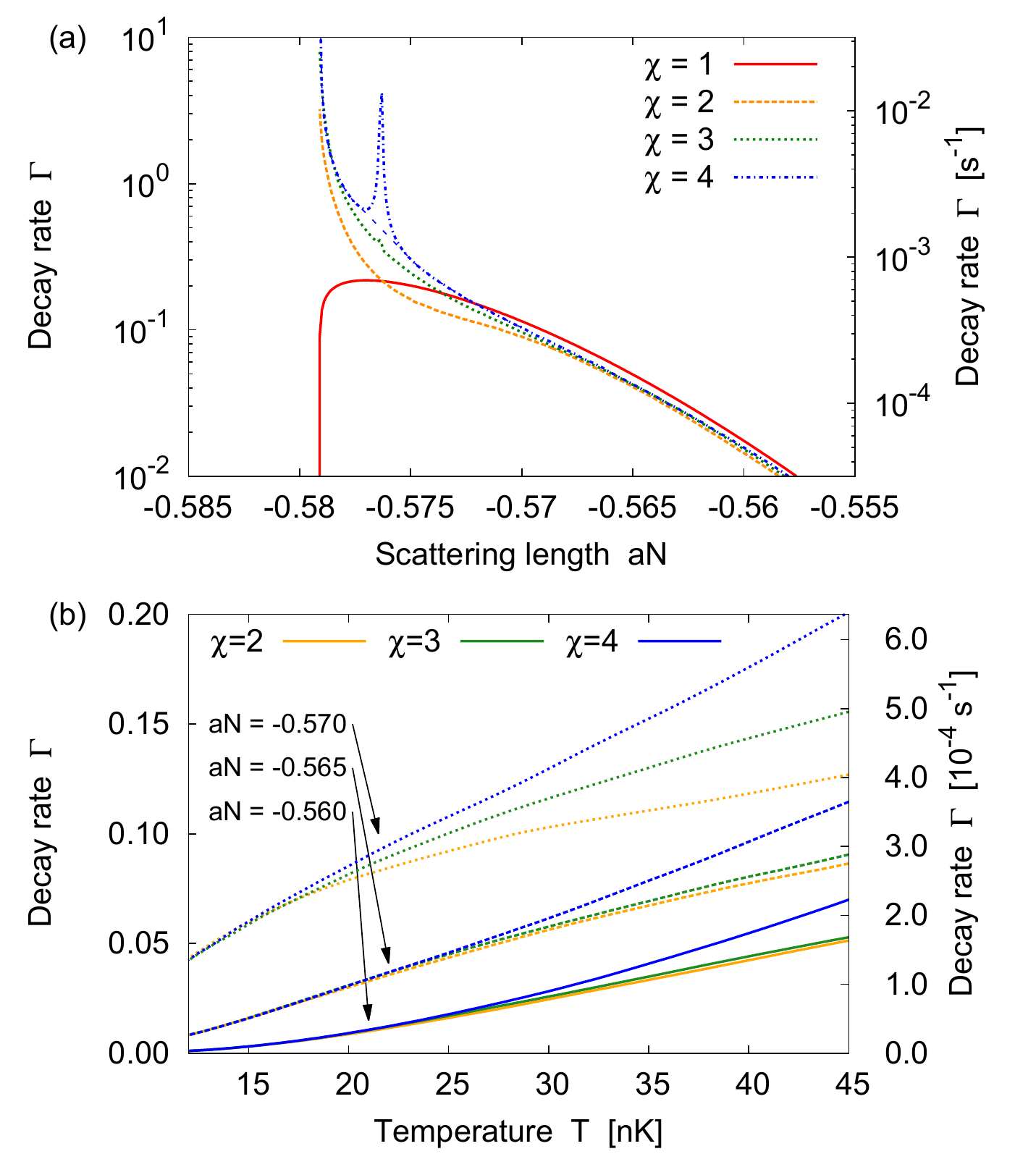}
\caption{
(Color online)
Decay rate of the thermally induced coherent collapse of a $^{87}$Rb BEC in a 
trap with frequency $\omega=2\pi\times100\,$Hz and at a temperature of 
$T=24\,$nK. The calculations have been performed for $\Ng=2$ coupled wave 
functions.
(a) 
The decay rate increases with decreasing scattering length and it reaches its 
highest values close to the critical value $\acrit N$. Also corrections of the 
higher-order normal form expansions become more important in this region.
(b)
The decay rate also increases with the temperature. Higher normal form orders 
predict higher decay rates throughout. At low temperatures, the corrections are 
small and already the second-order ($\chi=2$) approximation of the transition 
state can be sufficient. By contrast, higher normal form orders give significant 
corrections at higher temperatures.
}
\label{fig:alkali-decay-rate}
\end{figure}
%
Figure \ref{fig:alkali-decay-rate}(a) shows the decay rate in dependence on the 
scattering length at a temperature of $T=24\,$nK for $\Ng=2$ coupled Gaussians. 
With decreasing scattering length, the decay rate significantly increases. It 
has its largest values close to $\acrit N$ where the energy barrier is small.
The drop of the first-order rate ($\chi=1$) to zero at $\acrit N$ is not 
physical, since the barrier vanishes there and the rate should strongly 
increase. This tendency is correctly reproduced by the higher-order 
approximations ($\chi>1$) of the transition state.
Furthermore, it can be seen that corrections to the decay rate are significant  
near the critical scattering length, where better approximations of the 
transition state yield higher reaction rates.
We note that the peak in the fourth-order approximation at $aN \approx -0.5763$ 
is caused by the numerical resonance in the normal form coefficients that has 
already been discussed above [\cf\ \FIG~\ref{fig:alkali-NF-coef}(e)] and is not 
physical. The expected physical behavior is sketched by a dashed line.

We further note that Huepe \etal\ \cite{Huepe1999,Huepe2003} have also 
calculated the thermal decay rate of BECs with short-range interaction in the 
framework of a \emph{single} Gaussian trial wave function ($\Ng=1$). In the 
very vicinity of the critical scattering length or critical particle number, 
respectively, they obtained decay rates on the order of 
$10^{-2}$--$10^2\,$s$^{-1}$. By contrast, our results are on the order of 
$10^{-4}$--$10^{-2}\,$s$^{-1}$. 
Part of this deviation can be explained by the fact that they used different 
physical parameters for the trap frequency $\omega$ and the boson mass $m$. 
However, we expect that further significant corrections are caused by the 
different trial wave functions:
In the calculations of Huepe \etal\ the single Gaussian trial wave function  
only provides a single degree of freedom ($d=1$) which must be the reaction 
channel of the system. Since there are no further degrees of freedom a  
trajectory cannot avoid to undergo a reaction by leaving the reaction channel in 
a ``perpendicular'' direction. 
By contrast, we have $d=3$ for the trial wave function with $\Ng=2$ which
provides such additional degrees of freedom, so that a reaction becomes less 
probable.

In \FIG~\ref{fig:alkali-decay-rate}(b), the decay rate is presented in 
dependence of the condensate temperature $T$. 
Throughout, one finds that higher normal form approximations are less important 
at low temperatures and very important at higher temperatures.
Vice versa, the behavior of the decay rate in 
\FIG~\ref{fig:alkali-decay-rate}(b) can be used in order to estimate the 
temperature regime up to which a certain approximation will yield good results.

\subsection{BECs with long-range dipolar interaction}
\label{sec:dipolar}

Beyond BECs with short-range scattering interaction, also condensates with 
long-range and anisotropic DDI have been realized experimentally 
\cite{Griesmaier2005,Lu2011,Aikawa2012}. The investigation of the thermally 
induced coherent collapse in dipolar BECs has already been performed within the 
lowest-order approximation of the transition state 
\cite{Junginger2012d,Junginger2013b}. Therein, additional transition states 
emerged in bifurcations which gave rise to the expectation of a 
symmetry-breaking thermally induced collapse at certain physical parameters. 
It is not the scope of this paper to investigate this symmetry-breaking collapse 
scenario in more detail. Instead, we focus on the effects of higher-order normal 
form approximations of the transition state \cite{GelaBachelor}.

We therefore numerically solve the GPE \eqref{eq:GPE} with the variational 
ansatz \eqref{eq:TDVP-Gaussian-cyl} and apply the normal form procedure as 
described in \REF\ \cite{Junginger2014a}. We consider an external trap aspect 
ratio $\lambda=2$, where the dipolar BEC shows a conventional density 
distribution and where only a single transition state with cylindrical symmetry 
exists \cite{Junginger2013b}. 
The strength of the DDI in \EQ~\eqref{eq:potentials-DDI} is determined by the 
coefficient $\add N$ and it depends on the particle number. For an exemplary 
$^{52}$Cr condensate in a trap with frequency $\omega_\rho=2\pi \times 100\,$Hz, 
the parameter $\add$ has a value $\add \approx 0.0034$, and in case of 
$^{164}$Dy the value is $\add \approx 0.0536$.
For the following calculations, we use a model condensate with $\add N=1$ which 
represents each of these dipolar condensates with a respective particle number.

\begin{figure}[t]
\includegraphics[width=.49\columnwidth]{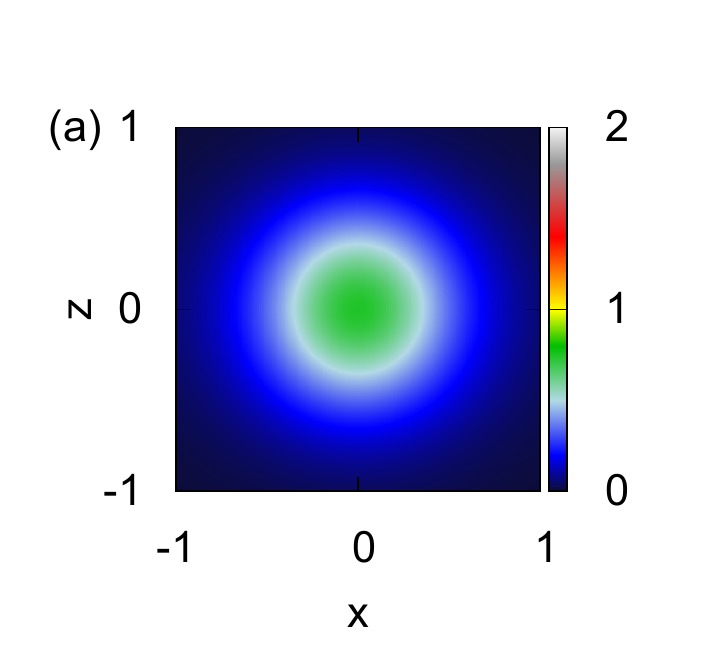} 
\includegraphics[width=.49\columnwidth]{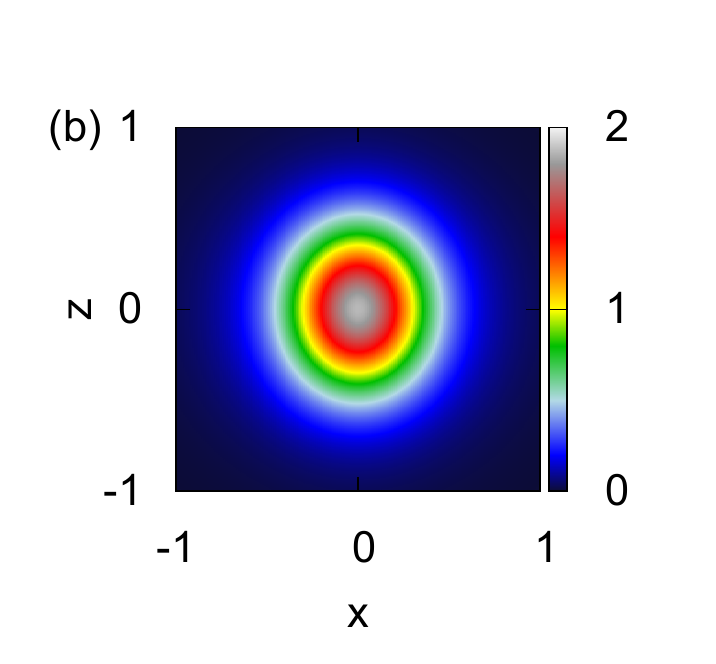}
\caption{
(Color online)
Density distribution $\abs{\psi(\rr)}^2$ (cut through the $x$-$z$-plane) of a 
dipolar BEC calculated with $\Ng=2$ coupled Gaussian wave functions at a 
scattering length $aN=-0.47$.
The density distribution of the ground state (a) is slightly extended more in 
the radial direction than in $z$-direction.
By contrast, the density distribution of the transition state (b) is more 
extended in $z$-direction. Moreover, it exhibits a highly increased density in 
the center of the trap as compared to the ground state.
}
\label{fig:dip-wave-functions}
\end{figure}

The physical meaning of the transition state in dipolar BECs is the same as in 
the case without long-range interaction as discussed in \SEC~\ref{sec:alkali}. 
However, because of the anisotropy of the DDI, the extensions of the atomic 
cloud in $\rho$- and $z$-direction differ (see 
\FIG~\ref{fig:dip-wave-functions}):
Due to the interplay between the external trap with the DDI which prefers an 
alignment of the dipoles in head-to-tail configuration, the ground state in 
\FIG~\ref{fig:dip-wave-functions}(a) only has a slightly larger extension in the 
radial than in $z$-direction. 
By contrast, the density distribution of the transition state in 
\FIG~\ref{fig:dip-wave-functions}(b) is more extended in $z$-direction and it 
exhibits a highly increased density in the center of the trap. Due to these two 
effects, the transition state corresponds to a highly attractive configuration 
of the interacting bosons.
As in the case without long-range interaction, this stationary state represents 
the condensate at the edge of the collapse. Any locally higher density and, with 
it, a higher attraction between the particles would lead to the collapse of the 
condensate.

\begin{figure}[t]
\includegraphics[width=\columnwidth]{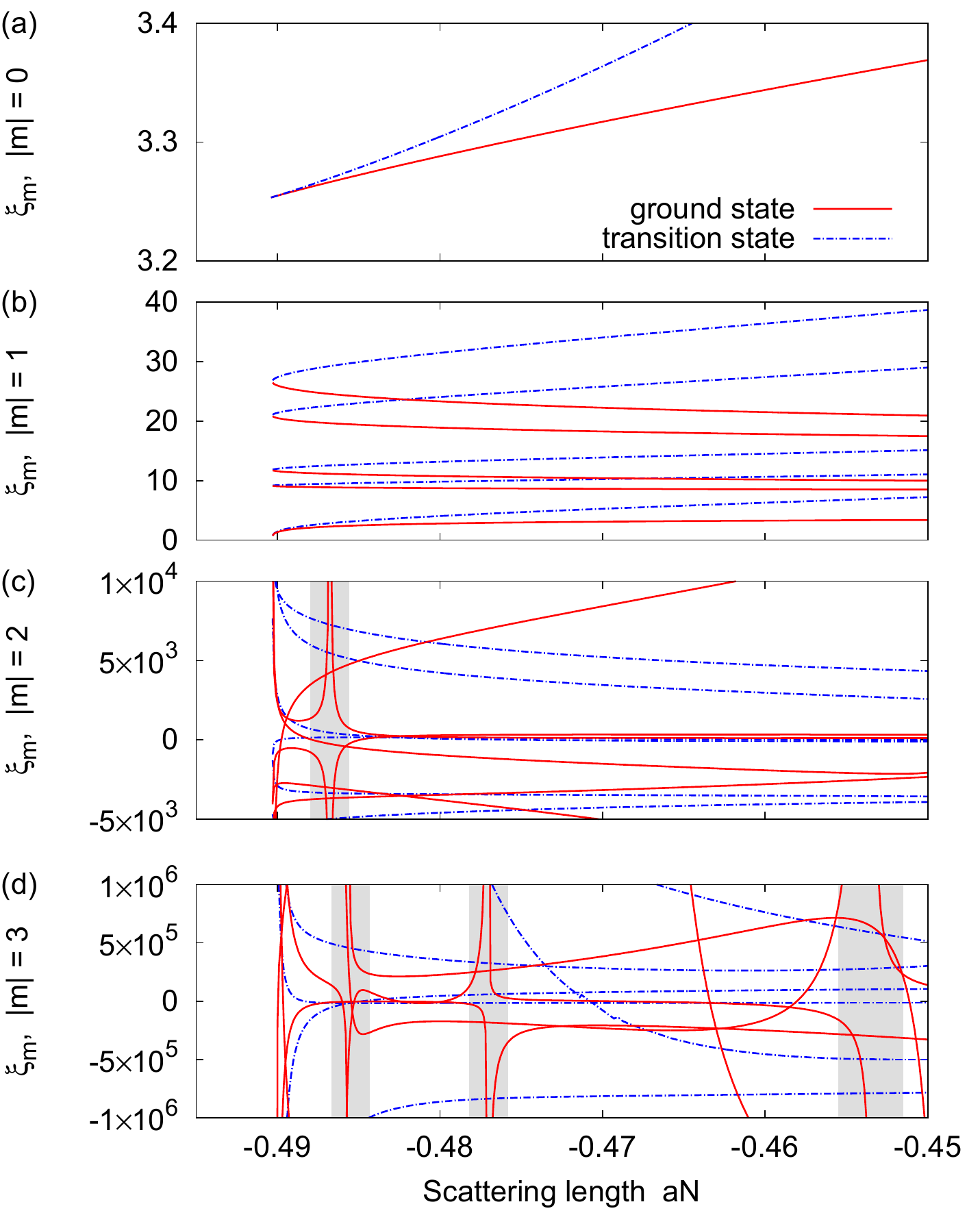}
\caption{%
(Color online)
Coefficients of the normal form expansion in different orders for a dipolar 
condensate with $\Ng=2$ coupled Gaussian wave functions.
As in \FIG~\ref{fig:alkali-NF-coef}, the fixed-point energies (a) of the ground 
and the transition state as well as the Bogoliubov eigenfrequencies (b) show a 
smooth behavior and the respective values merge at the critical scattering 
length $\acrit N=-0.4903$.
The coupling coefficients (c),(d) of the collective oscillation modes show poles 
at different values of the scattering length $a$ (highlighted by the gray bars), 
whose number increases with higher normal form orders. (Again, only a selection 
of the coefficients with a typical behavior is presented for the sake of 
clarity).
}
\label{fig:dip-NF-coef}
\end{figure}

Again, the full information about the condensate's local dynamics is reflected 
by the normal form coefficients which are shown in \FIG~\ref{fig:dip-NF-coef}, 
and also the interpretation of the individual orders of the expansion is the 
same as in \SEC~\ref{sec:alkali}.
The energy eigenvalues [see \FIG~\ref{fig:dip-NF-coef}(a)] and the frequencies 
of elementary excitations [see \FIG~\ref{fig:dip-NF-coef}(b)] show a smooth 
dependence on the scattering length and they merge at the latter's critical 
value.
Also, we observe that the coupling terms of the modes [see a selection in 
\FIGS~\ref{fig:dip-NF-coef}(c),(d)] are important because they have numerically 
large values.
What is different is that a resonance can already be found in the second order 
$\abs{\mm}=2$ where some coefficients diverge (highlighted by a gray bar in the 
background of the plot). The reason for this earlier occurrence of resonances 
compared to that in \FIG~\ref{fig:alkali-NF-coef} is, here, that the variational 
approach \eqref{eq:TDVP-Gaussian-cyl} exhibits more degrees of freedom than the 
ansatz \eqref{eq:TDVP-Gaussian-radial}. As a consequence, there exist more 
possible oscillation modes that can couple according to 
\EQ~\eqref{eq:rational-independent}, which also leads to three more resonances 
in the third normal form order in \FIG~\ref{fig:dip-NF-coef}(c). 
Analogously to the case discussed in \FIG~\ref{fig:alk-resonance}, also each of 
these resonances can be identified with a certain mode coupling of higher 
harmonics in the dipolar BEC. However, a detailed investigation of these mode 
couplings goes beyond the scope of this paper and we refer the reader to \REF\ 
\cite{GelaBachelor} for further studies including the effect of varying external 
parameters on the resonances.

Furthermore, we note that a detailed comparison between the exact Bogoliubov 
eigenfrequencies of dipolar BECs with those frequencies obtained from the 
variational approach using coupled Gaussian wave functions can be found in \REF\ 
\cite{Kreibich2013a}. Therein, it is shown that, as in the case of BECs without 
long-range interaction, the lower Bogoliubov frequencies of dipolar BECs can be 
well reproduced by the variational approach, while the convergence becomes worse 
for higher-frequency modes. 
Again, a check of the normal form coefficients in 
\FIG~\ref{fig:alkali-NF-coef}(c)--(d) would be possible by numerically 
evaluating higher-order coupling coefficients of the Bogoliubov modes.

\begin{figure}[t]
\includegraphics[width=\columnwidth]{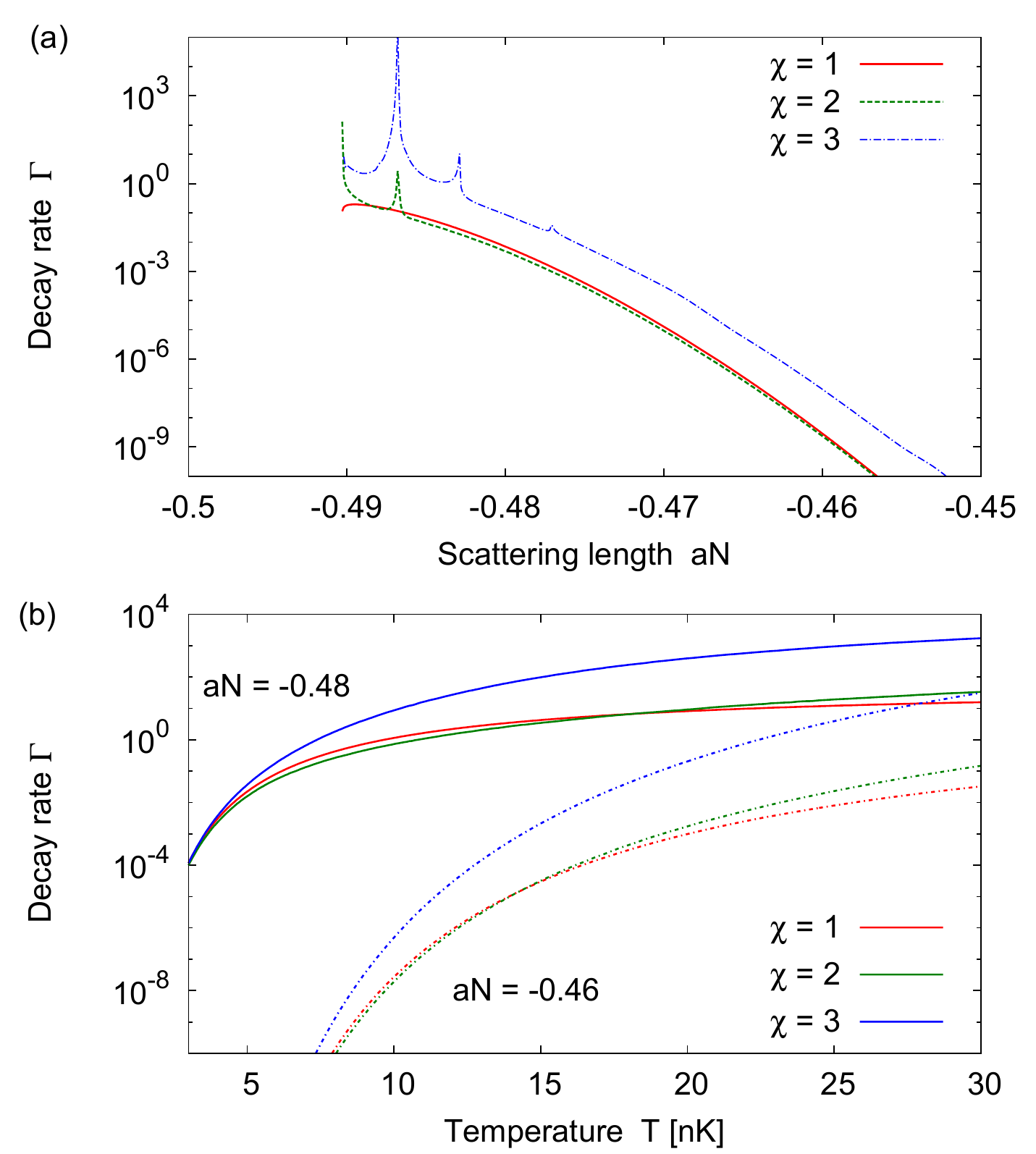}
\caption{
(Color online)
(a) Thermal decay rate of a dipolar BEC in dependence on the scattering length 
and for different normal form orders $\chi$ (parameters: $\beta=200$, $\Ng=2$).
(b) Decay rate in dependence on the temperature.
}
\label{fig:dip-decay-rate}
\end{figure}

In \FIG~\ref{fig:dip-decay-rate}, the thermal decay rate is shown in dependence 
on the scattering length and the temperature $T$.
According to \FIG~\ref{fig:dip-decay-rate}(a) higher-order corrections are, 
again, important: The second-order decay rate ($\chi=2$) gives only small 
corrections to the decay rate compared to the first order ($\chi=1$). However, 
importantly, the second order correctly reproduces an increasing decay rate when 
one approaches the critical scattering length. 
For the given temperature, the third-order normal form predicts an increased 
reaction rate over the whole range of the scattering length of about one order 
of magnitude.
Again, we note that the single peaks are a consequence of the resonances shown 
in \FIG~\ref{fig:dip-NF-coef} and that they are not physical.
As presented in \FIG~\ref{fig:dip-decay-rate}(b), the higher-order corrections 
to the decay rate are again important at high temperatures, whereas the 
first-order approximation of the transition state already is appropriate for 
small temperatures.

\section{Conclusion and outlook}

In this paper, we have investigated the properties of the ground and transition 
state in BECs with short- and long-range interaction. 
In both systems, we have discussed the transition state as a certain density 
distribution of the atomic cloud that typically exhibits a locally increased 
density in the center of the trap as compared to the ground state.
Higher-order normal form approximations to the local dynamics of the condensate 
in the vicinity of their fixed points turned out to be important throughout, 
because their large numerical values induce significant corrections already at 
small deviations from the stationary states.
As a general tendency, we observed that low-order normal form contributions 
converge quite fast when the trial wave function is improved, while higher-order 
corrections show slower convergence.

Calculating the decay rate of the condensates by applying TST, we observed 
significant corrections especially close to the critical value of the scattering 
length where the attraction dominates in the system. 
Higher-order normal form approximations are capable of reproducing the 
physically expected behavior of a monotonically increasing reaction rate when 
the critical scattering length is approached. In general, the reaction rates 
within higher-order approximations are dramatically increased compared to the 
usual harmonic approximation of the transition state, which gives rise to the 
expectation that the decay mechanism of the thermally induced coherent collapse 
can play an even more important role than estimated in previous investigations 
\cite{Junginger2013b}.

Finally, our investigations revealed that resonances in the normal form 
procedure become more and more important in higher-order approximations and for 
a large number of degrees of freedom. 
To appropriately treat these resonances, 
the normal form expansions presented in \REF\ \cite{Junginger2014a} need to be 
adapted to this situation which is currently work in progress.
Another way to circumvent divergences in the normal form coefficients could be 
the application of uniform approximations \cite{Junginger2012d} to the 
Hamiltonian by which bifurcations in the transition state have already been 
handled successfully.

The BECs in this paper have been considered in spherically or cylindrically 
symmetrical external traps. However, it is emphasized that the normal form 
procedure is not limited to this case, but also external traps with tri-axial 
symmetry can be treated (\cf\ \REFS\ \cite{Junginger2012d,Junginger2013b}). The 
higher number of degrees of freedom, however, significantly increases the 
numerical effort.
It will be a future task to investigate the influence of additional bath-degrees 
of freedom to the reaction rate.
As already mentioned above, another task will be to compare the numerical 
values of the normal form coefficients with numerical results from high-order 
expansions of the GPE.
We hope to induce further experiments with 
our manuscript performing detailed measurements of the thermal decay rate close 
to the critical scattering length for BECs with short- and long-range 
interaction.

\section*{Author contribution statement}

A.~Junginger and J.~Main have developed and implemented the theory, T.~Schaller 
and G.~H\"ammerling have calculated the results for BECs with short- and 
long-range interaction. A.~Junginger wrote the manuscript, but all authors have 
been involved in its preparations.


\end{document}